\documentclass[reprint,aps,prl,longbibliography,superscriptaddress]{revtex4-2}

\usepackage{graphicx}       
\usepackage{hyperref}       
\usepackage{amsmath,amssymb,amsfonts}  
\usepackage{multirow}       
\usepackage{xcolor}         
\usepackage{booktabs}       
\usepackage{float}          

\graphicspath{{figs/}} 

\begin{document}

\author{Yaprak Önder}
    \affiliation{Max Planck Institute for the Physics of Complex Systems, 01187 Dresden, Germany}
    \affiliation{Department of Physics, Harvard University, Cambridge MA, USA, 02139}
\author{Abbas Ali Saberi}
\email{corresponding author: asaberi@constructor.university}
\affiliation{School of Science, Constructor University, Campus Ring 1, 28759 Bremen,
Germany}
\affiliation{Max Planck Institute for the Physics of Complex Systems, 01187 Dresden, Germany}
 
\author{Roderich Moessner}
        \affiliation{Max Planck Institute for the Physics of Complex Systems, 01187 Dresden, Germany}

\title{$q$-Gaussian Crossover in Overlap Spectra towards 3D Edwards–Anderson Criticality}

\date{\today}

\begin{abstract}
We introduce a spectral approach to characterizing the three-dimensional Edwards-Anderson spin glass. By analyzing the eigenvalue statistics of overlap matrices constructed from two-dimensional cross-sections, we identify a crossover from the Wigner semicircle law at high temperatures towards a Gaussian distribution, {which is consistently attained near the spin-glass critical point.} Visible for different distributions of the random coupling, the Gaussian distribution can potentially serve as a {robust} spectral indicator of criticality. Remarkably, the spectral density is well-described by Tsallis statistics, with the entropic index $q$ evolving from $q = -1$ (semicircle, $T=\infty$) to $q = 1$ (Gaussian) at $T_c$, revealing a statistical structure inside the paramagnetic phase. We find $q\le 1$ within numerical precision. While the local level statistics remain consistent with GOE statistics, reflecting standard level repulsion, the temperature dependence appears mainly in the global spectral density. Our results present spectral statistics as a computationally efficient complement to multi-replica correlator methods and provide a new perspective on cooperative and critical phenomena in disordered systems.
\end{abstract}

\maketitle

\textit{Introduction.}---Spin glasses, paradigmatic disordered systems with frustration, exhibit phase transitions governed by intricate energy landscapes and nonergodic dynamics \cite{MezardParisiVirasoro, FischerHertz}. In three dimensions, the Edwards-Anderson (EA) spin glass \cite{Edwards1975} is a well-studied model. While significant progress has been made in characterizing the properties of equilibrium and critical exponents \cite{Katzgraber2006},  the nature of its critical behavior remains incompletely understood. Competing theories, such as replica symmetry breaking (RSB) \cite{Parisi1983} and the droplet picture \cite{FisherHuse1986}, offer descriptions of the freezing transition, but reconciling their predictions with numerical and experimental signatures remains a challenge \cite{moore2021droplet}.

One possible avenue towards understanding spin glass criticality lies in characterizing its properties beyond conventional order parameters. Spectral methods, particularly those rooted in random matrix theory (RMT), have turned out to be powerful tools for uncovering universal statistical properties in complex systems ranging from quantum chaos to high-dimensional disordered materials \cite{Mehta1991, Guhr1998, Forrester2010}. In statistical physics, RMT has captured critical signatures in systems such as the Anderson metal-insulator transition \cite{Evers2008} and percolation models \cite{Saber2022, Malekan2022}, while recent studies have demonstrated distinct spectral characteristics at criticality in two-dimensional Ising models with local and nonlocal interactions \cite{saberi2024interaction}. However, the potential of RMT to reveal properties of spin glasses remains largely unexplored. Unlike earlier spin/field-based matrix constructions \cite{saberi2024interaction}, we use an overlap-based matrix ensemble for the 3D EA spin glass, which lets us track how the bulk spectrum changes across the critical region. Given the intricate correlations governing the spin glass phase, it provides a promising probe of criticality and fluctuations in the presence of disorder.

Signatures of 3D criticality can leave robust imprints on lower-dimensional subsystems~\cite{dashti2015statistical}. In ordered magnets, for example, the 2D cross-sections of the 3D Ising model exhibit percolation transitions at the bulk Curie temperature~\cite{Saberi2010, arenzon2015slicing}, while the interfacial roughness diverges with the system size at criticality~\cite{dashti2019two, rodriguez2025anomalous}.
Reducing the system to two dimensions is attractive as it provides access to a wealth of powerful analytical and computational tools, including RMT, conformal field theory (CFT), and stochastic Loewner evolution (SLE)~\cite{Saberi2010}.

Here, we use this dimensional reduction to explore the physics of spin glasses: we construct overlap matrices from 2D slices of the 3D EA spin glass and analyze their spectral statistics. We identify and characterize their temperature dependence as correlations build up and criticality sets in, providing an RMT-rooted signature of the spin glass transition without the need for direct computation of multi-replica correlators~\cite{Beenakker1997}.
Specifically, we explore the spectral statistics from the high-temperature regime down to and, for the smaller systems, slightly below the critical temperature $T_c=1/\beta_c$. 
While at infinite temperature, the spectral properties are expected to follow standard RMT, we study how disorder and frustration shape the spectrum as the temperature is varied. We emphasize that the overlap matrix analyzed below is not a quantum Hamiltonian and its eigenvalues are not energy levels; rather, it is a classical object constructed from inter-replica overlaps on a cross-section, and the spectral analysis is used as a statistical fingerprint of correlations and critical fluctuations.

We find that the spectral density crosses over from the Wigner semicircle law at very high temperature to a near-Gaussian form when approaching criticality. Remarkably, the bulk spectral density at intermediate temperatures is well-described by $q$-Gaussian distributions, with the Tsallis parameter $q$ evolving from $q \to -1$ (semicircle) to $q = 1$ (Gaussian) at criticality. Our findings are similar for different disorder distributions (Gaussian or bimodal), {suggesting a robust, detail-independent evolution between distinct spectral behaviors leading up to, and at, the critical point.} These findings establish spectral statistics as a computationally efficient framework for investigating spin glass properties and criticality \footnote{{By “computationally efficient” we mean efficiency post equilibration in two concrete senses: (i) per-sample cost and memory---one dense eigendecomposition of an $L\times L$ real–symmetric matrix ($O(L^{3})$ floating-point operations for standard dense tridiagonalization/QR or divide-and-conquer routines); (ii) each sample yields $L$ bulk eigenvalues, so at fixed binning the RMS error of the spectral histogram (and of $D_{\rm KL}$ or the $q$-fit) scales as $[L\,N_{\rm samp}]^{-1/2}$, i.e. $N_{\rm samp}=O(1/(L\varepsilon^{2}))$ to reach accuracy $\varepsilon$. As with standard approaches, equilibration is the dominant cost and finite-size scaling $L\to\infty$ is required.}}. Their use as general indicators of critical behavior in disordered systems remains a tantalizing direction for future exploration.

\textit{Model.}---  
We study the 3D Edwards-Anderson (EA) Ising spin glass, described by the Hamiltonian:  
\begin{equation}
    H = -\sum_{\langle \mathbf{x,y} \rangle} J_{\mathbf{xy}} s_\mathbf{x} s_\mathbf{y}
\end{equation}
where $s_{\mathbf{x}} = \pm 1$ denotes an Ising spin at site $\mathbf{x} = (i,j,k)$ on a cubic lattice, and $\langle \mathbf{x,y} \rangle$ represents nearest-neighbor pairs. The coupling constants $J_{\mathbf{xy}}$ are drawn either from a Gaussian distribution with zero mean and unit variance or from a bimodal distribution $J_{\mathbf{xy}} = \pm J$ with equal probability. {Periodic boundary conditions are applied in all spatial directions.}

To analyze spectral properties, we construct overlap matrices from two independent replicas of the 3D EA model. {These replicas are generated for the {\it same} disorder realisation and temperature, but in independent Monte Carlo runs.} We consider the 2D cross-section at $k = L/2$ of the respective spin configurations $s^{(1)}_{ij}$ and $s^{(2)}_{ij}$ at sites $(i,j)$ in the plane. With periodic boundary conditions and i.i.d.\ couplings, the disorder distribution (and hence the disorder-averaged ensemble) is translationally invariant, so any fixed-$k$ plane is statistically equivalent; choosing $k=L/2$ is therefore only a convention. We define the overlap matrix  as $\mathcal{M} = \frac{1}{2} \left( \mathcal{M}' + \mathcal{M}'^T \right)$, where the local spin overlap is given by  
\begin{equation}\label{Eq-2}
    \mathcal{M}'_{ij} =  
    s^{(1)}_{ij} s^{(2)}_{ij},
\end{equation}
with $(\cdot)^T$ denoting the matrix transpose. The matrix $\mathcal{M}$ is symmetrised to ensure a real spectrum. Its off-diagonal elements take values  $\{\pm 1, 0\}$, while diagonal elements are $\pm 1$. To ensure consistency with standard random matrix theory (RMT) in the limit $T \to \infty$, we rescale $\mathcal{M}$ by $1/\sqrt{L}$ so that the bulk eigenvalues lie within the interval $[-\sqrt{2}, \sqrt{2}]$.

To generate equilibrated spin glass configurations, we employ a combination of Monte Carlo sweeps, Houdayer cluster updates, and parallel tempering (exchange Monte Carlo) moves, following the algorithmic approach detailed in \cite{Zhu2015}. Equilibration is verified using the energy–overlap consistency relation established for short-range models with Gaussian disorder, $[\langle e \rangle]_J = \beta \left( [\langle q_l \rangle]_J - 1 \right)$ \cite{Equilib}, where $e$ is the energy per bond and $q_l$ is the link overlap defined as $q_l = \frac{1}{N_b} \sum_{\langle \mathbf{x,y} \rangle} s^{(1)}_{\mathbf{x}} s^{(1)}_{\mathbf{y}} \cdot s^{(2)}_{\mathbf{x}} s^{(2)}_{\mathbf{y}}$, with $N_b$ denoting the total number of bonds. Here, $[\langle \cdot \rangle]_J$ represents both the thermal average and the average over disorder realizations. {The criterion for equilibration is for both sides of the consistency relation to agree within statistical uncertainty.} Our temperature range spans the regime between $\beta = 0.1$ (i.e., $T \gg T_c$) down to $\beta \approx \beta_c$. Due to the severe dynamic slowdown near criticality~\cite{10.3389/fphy.2025.1563982,Nakamura_2019}, the simulations are limited to small system sizes near criticality. For system sizes up to $L=22$, we equilibrate samples down to the critical temperature; for larger system sizes, the data sets are cut at the coldest temperature at which the equilibration signal is reached.

\textit{Spectral crossover of bulk eigenvalues.}---
Our central object of study is the spectral density of eigenvalues from ensembles of overlap matrices $\mathcal{M}$ constructed at various temperatures. At very high temperatures, the matrix entries are approximately independent and have zero mean, resulting in a spectral density that follows the Wigner semicircle distribution. As $T$ decreases, correlations begin to emerge, leading to a nonzero mean that causes a single eigenvalue to detach from the bulk~\footnote{For large $L \times L$ real-symmetric random matrices with entries of mean $\mu/L$ and variance $\sigma^2/L$, an eigenvalue detaches from the Wigner semicircle when the mean exceeds the disorder scale, i.e., $\lvert \mu \rvert > \sigma$ \cite{S_F_Edwards_1976}. In this regime, the largest eigenvalue converges to $\lambda_{\rm max} \approx \lvert \mu \rvert + \sigma^2 / \lvert \mu \rvert$. This result has been extended to matrices with absolutely summable (i.e., sufficiently short-range) correlations, where the outlier eigenvalue exhibits Gaussian fluctuations \cite{chakrabarty2024largest}.}. Since our focus is on the bulk eigenvalue statistics, we exclude this outlier from our analysis.

\begin{figure}[t]
    \centering    \includegraphics[width=1.12\linewidth, trim=0 0 0 0, clip]{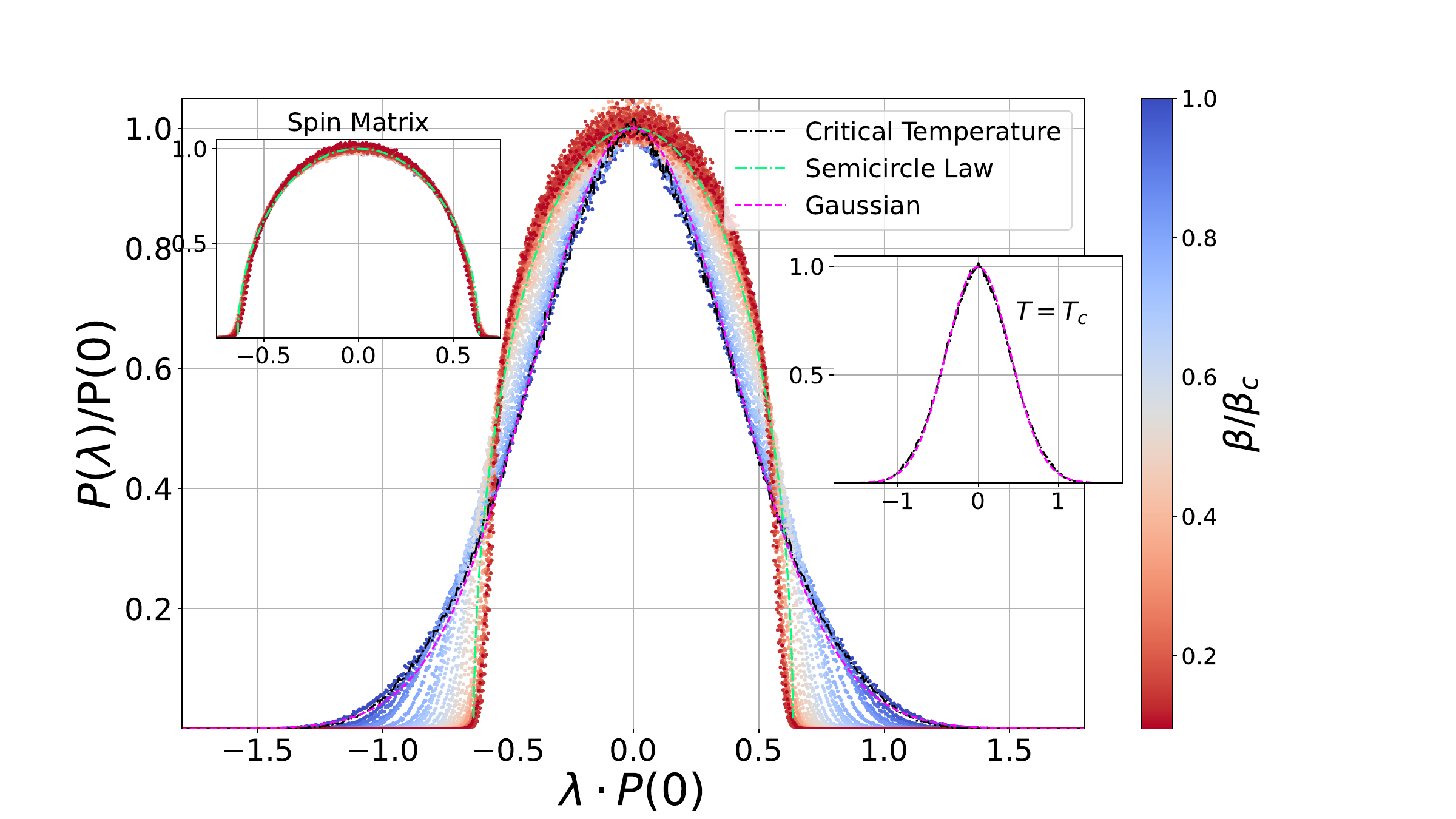}
    \caption{Scaled spectral density $P(\lambda)/P(0)$ of bulk eigenvalues for the overlap matrix $\mathcal{M}$, shown across temperatures for system size $L = 22$. Spectra are rescaled along both axes by $P(0)$ to account for the peak height and the temperature-dependent spectral width, enabling direct shape comparison across temperatures. A smooth crossover is observed from the Wigner semicircle (dashed green) at high temperatures to a Gaussian form (dashed magenta) near the critical point $\beta_c \simeq 1.05$ (i.e., $T_c \simeq 0.952$; black dashed line). Right inset illustrates the close agreement of the spectral shape with a Gaussian at $\beta=\beta_c$. Left inset: the matrix built from a single replica, $\mathcal{M}^{\prime s}_{ij}=s_{ij}$, yields a semicircle at all $T$. 
    }
    \label{fig:center22}
\end{figure}

Figure~\ref{fig:center22} shows the normalized spectral density of the overlap matrix $\mathcal{M}$, obtained from thermally equilibrated configurations of the 3D EA model across a range of temperatures, parametrized by $\beta = 1/T$ (color scale). At high temperatures (low $\beta$), the spectrum closely follows the Wigner semicircle law (green dashed line), consistent with standard RMT. As $\beta$ increases, the spectrum gradually deviates from the semicircle and approaches a Gaussian shape, reaching close agreement at the critical point $\beta_c \simeq 1.05$ (i.e., $T_c \simeq 0.952$; dashed magenta). The inset highlights this match at $T_c$, confirming the onset of Gaussian statistics at criticality.

The robustness of this phenomenon is further supported by the observation of a similar spectral evolution for bimodal random couplings, $J_{\mathbf{xy}} = \pm J$ (see Fig.~\ref{fig:center30pm}). {The coincidence of the Gaussian form with the critical temperature in both disorder types suggests that this spectral crossover provides a robust and detail-insensitive indicator of spin-glass criticality.}

\begin{figure}[t]
    \centering
    \includegraphics[width=\linewidth , clip]{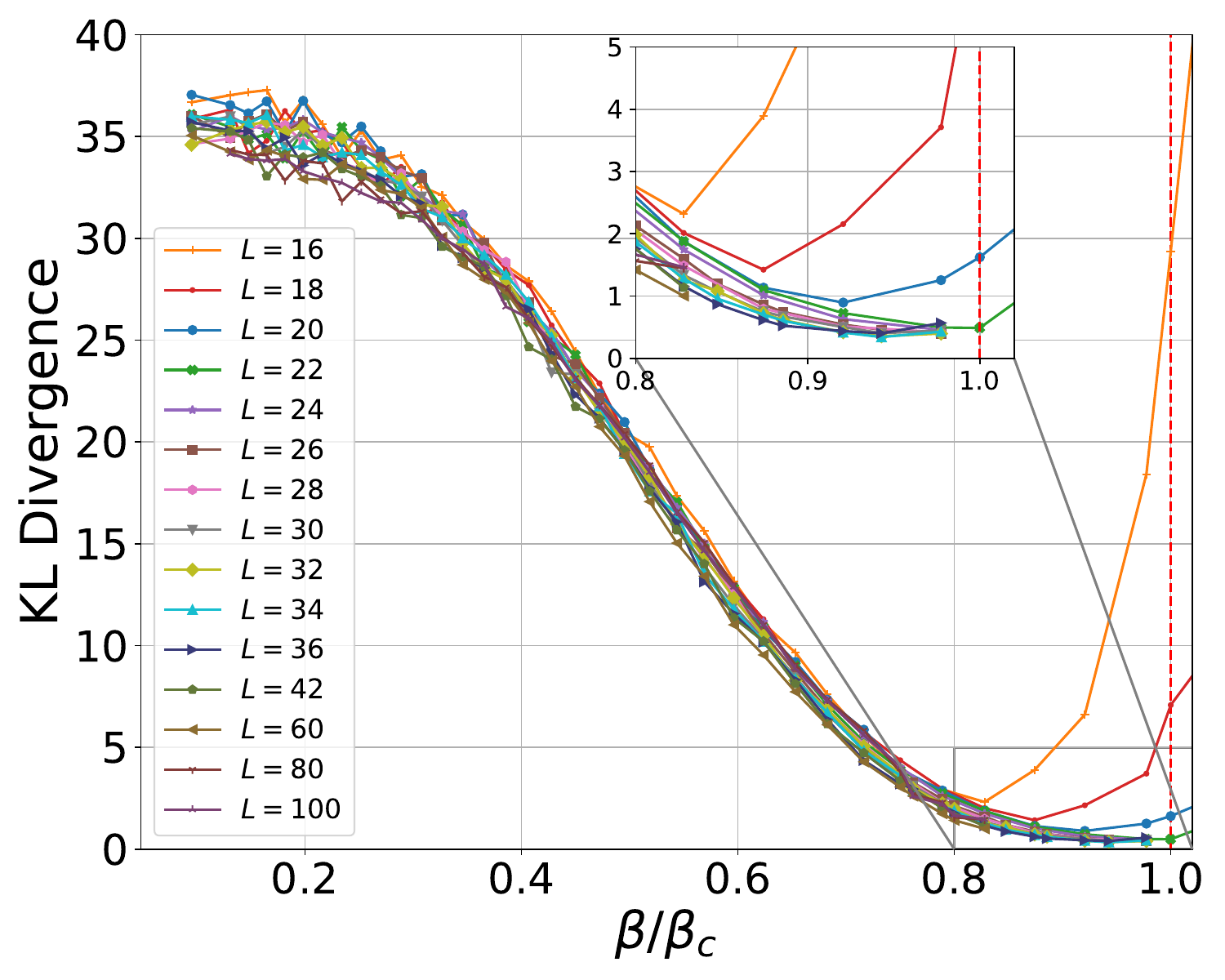}
    \caption{Temperature dependence of $D_{\rm KL}$ between the numerically obtained bulk spectrum and a variance-matched Gaussian for system sizes $L=16$ to $L=100$ (for large $L$, data are shown over a restricted temperature range). For $\beta/\beta_c \gtrsim 1$ (vertical red dashed line), $D_{\rm KL}$ decreases with $L$ and tends toward zero, indicating convergence of the bulk spectral density to a Gaussian.}
    \label{fig:Fig3b}
\end{figure}

To quantify the spectral crossover from the Wigner semicircle law to a Gaussian distribution in two-dimensional cross-sections of the 3D EA model, we compute the Kullback–Leibler (KL) divergence between the empirical spectral density $P(\lambda)$ at each temperature and a reference Gaussian distribution $P_{\rm G}(\lambda)$ (see Fig.~\ref{fig:Fig3b}):
\begin{equation}
    D_{\rm KL}(P \parallel P_{\rm G}) = \int P(\lambda) \ln \frac{P(\lambda)}{P_{\rm G}(\lambda)} \, d\lambda.
\end{equation}
Since the spectral densities near $T_c$ exhibit a symmetric, bell-shaped profile, the key differences arise in the behavior of the tails. The logarithmic weighting in $D_{\rm KL}$ amplifies these deviations, making it a particularly sensitive probe of the crossover. To avoid over-reliance on KL sensitivity to tails, we also confirmed Gaussianity near $T_c$ using an integrated squared difference $D_{L^2}=\int [P(\lambda)-P_G(\lambda)]^2\, d\lambda$ to a variance-matched Gaussian and excess-kurtosis trends.

{As shown in Fig.~\ref{fig:Fig3b}, $D_{\rm KL}$ is highest at high temperatures ($\beta \lesssim 0.3$), where the spectral density follows the Wigner semicircle law, and decreases progressively as $\beta$ increases. Near the critical point $\beta_c \simeq 1.05$, $D_{\rm KL}$ approaches zero, indicating a Gaussian distribution, i.e., $P(\lambda) \approx P_{\rm G}(\lambda)$. For $\beta > \beta_c$, finite-size effects become more pronounced: $D_{\rm KL}$ decreases systematically with increasing system size, suggesting that the bulk spectral density remains close to Gaussian for $\beta \gtrsim \beta_c$ and becomes increasingly consistent with a Gaussian as $L$ increases. We note that the behavior for the bimodal disorder distribution is largely similar }(see Fig.~\ref{fig:kldiv_pm}).

\begin{figure*}[t]
    \centering
    \includegraphics[width=1.0\linewidth , clip]{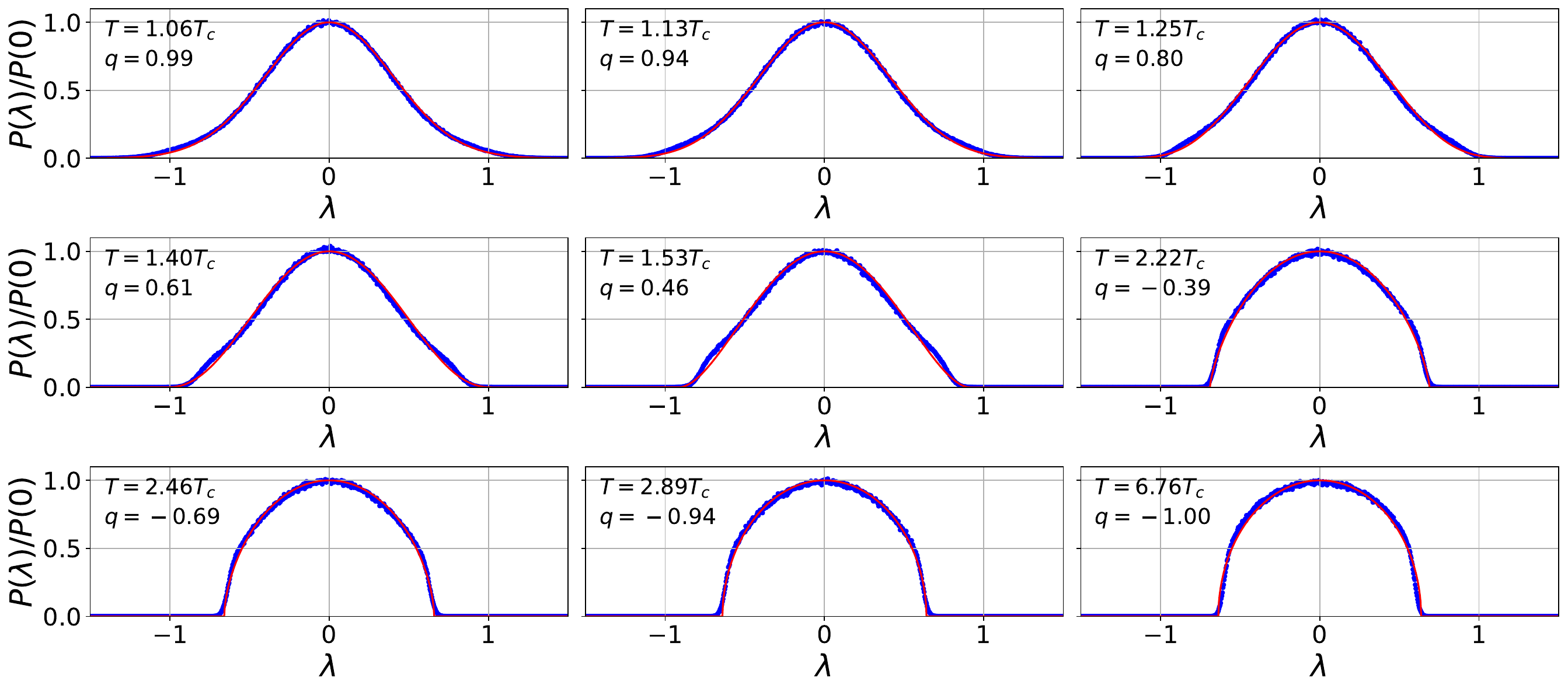}
    \caption{
    Bulk eigenvalue distributions $P(\lambda)$ for system size $L = 36$ at selected temperatures,  empirically fitted to the $q$-Gaussian form \eqref{eqn:qgaussian} (red curves). At high temperatures, the spectral density is close to the Wigner semicircle law ($q = -1$), and  smoothly crosses over to a Gaussian distribution ($q = 1$) at criticality $T \approx T_c$ (i.e., $\beta \approx \beta_c$).
    }
    \label{fig:Fig4}
\end{figure*}
\textit{$q$-Gaussian crossover in the paramagnetic phase.}---
On top of quantifying the approach to Gaussianity via $D_\mathrm{KL}$, we have found a one-parameter family of distributions which captures the spectral crossover quantitatively: we observe that the bulk eigenvalue distribution is well-described by a $q$-Gaussian, a family of distributions that arises  in the framework of nonextensive statistical mechanics \cite{tsallis2009introduction, tsallis_2009}. The $q$-Gaussian form has been employed in diverse contexts involving correlated, non-Markovian, and constrained systems. It interpolates between compact-support shapes for $q<1$ (including the semicircle at $q=-1$), the ordinary Gaussian at $q=1$, and power-law (heavy) tails only for $q>1$ \cite{tsallis2009introduction, tsallis_2009}. In our data the fitted values satisfy $q\le 1$ for all temperatures and sizes studied, i.e., we do not observe power-law (heavy) tails.

The bulk eigenvalue distributions at representative temperatures are empirically fitted (red curves in Fig.~\ref{fig:Fig4}) to the $q$-Gaussian form:
\begin{equation}
    P(\lambda) = \frac{C_q}{\lambda_0}\left[1 - (1 - q) \left(\frac{\lambda}{\lambda_0}\right)^2\right]^{\frac{1}{1 - q}},
    \label{eqn:qgaussian}
\end{equation}
where $C_q$ is a normalization constant and $\lambda_0$ is a scale parameter.

\begin{figure}[b]
    \centering
    \includegraphics[width=\linewidth , clip]{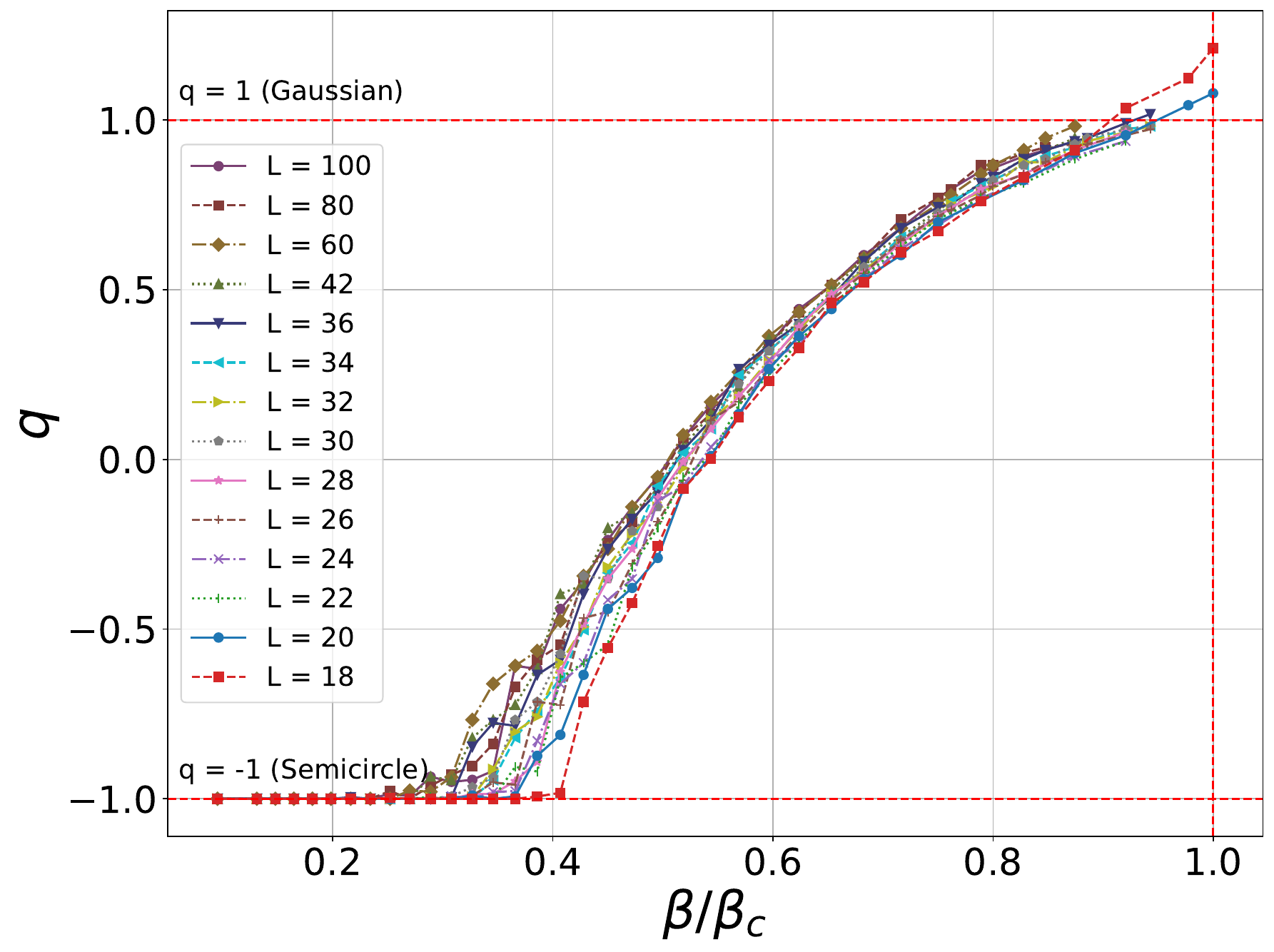}
    \caption{{Temperature dependence of the Tsallis entropic index $q$, extracted by minimizing the KL divergence between the empirical spectral density and the $q$-Gaussian fit, Eq. \ref{eqn:qgaussian}, for system sizes $L = 18$ to $L = 100$. The curves begin to depart from the semicircle value $q = -1$ at progressively lower $\beta$ for larger systems.}
    }
    \label{fig:Fig5}
\end{figure}

The optimal $q$ value at each temperature is obtained by minimizing the KL divergence between the spectral density and the $q$-Gaussian distribution. Figure~\ref{fig:Fig5} shows the temperature dependence of $q$ for different system sizes $L$. 

The function $q(T)$ thus effectively parametrizes the crossover from the semicircle law ($q=-1$) at high temperatures to Gaussian behavior ($q=1$) near criticality and its persistence below $T_c$. The inset of Fig.~\ref{fig:Fig5} highlights the region near $T_c$, showing that $q$ reaches unity within numerical precision.

\textit{Discussion and Conclusions.}
Our study reveals that the spectral statistics of overlap matrices constructed from two-dimensional cross-sections of the 3D EA spin glass model evolve nontrivially  as temperature decreases. The onset of $q$-Gaussian spectral statistics well above $T_c$ suggests that even the paramagnetic phase may harbor  internal structure. It is important to distinguish between two sources of disorder in the system: the quenched randomness of the coupling constants, and the thermal fluctuations of spin configurations within a single replica. Our numerical tests confirm that matrices constructed directly from the spin configurations of a single replica, by assigning
\begin{equation}
    \mathcal{M}_{ij}^{'s} = s_{ij} \label{eq:msprime}\ ,
\end{equation}
and symmetrizing/rescaling as above yield spectral densities that follow the semicircle law across all temperatures (see the left Inset in Fig. \ref{fig:center22}). In contrast, the  spectral deformation arises specifically in the overlap between two replicas,  indicating that the relevant structure is not encoded in individual configurations, but rather in their inter-replica correlations.

In various complex systems, $q$-Gaussians have been associated with long-range correlations, memory effects, and hierarchical structure in phase space \cite{tsallis2009introduction, wilk2000interpretation}. In this context, the observed spectral crossover may reflect the progressive buildup of correlated fluctuations, even before the onset of thermodynamic criticality.

It is important to emphasize that, unlike microscopic eigenvalue statistics such as level spacing distributions, which are governed by universal laws in random matrix theory, the global spectral density \footnote{This refers to the disorder- and thermal-ensemble averaged bulk density obtained by pooling bulk eigenvalues over disorder realizations and Monte Carlo samples.} is a non-universal quantity and thus sensitive to details of the underlying correlations. In particular, even relatively weak two-point correlations among matrix elements can significantly reshape the global spectral form, while leaving local eigenvalue statistics largely unaffected. 

We have checked this explicitly by computing the adjacent-gap ratio $r=\left\langle \min(\Delta_i,\Delta_{i+1})/\max(\Delta_i,\Delta_{i+1})\right\rangle$, where $\Delta_i=\lambda_{i+1}-\lambda_i$ are consecutive eigenvalue gaps in the bulk: within statistical accuracy it shows clear level repulsion, no systematic temperature dependence across the range studied, and a drift toward the GOE reference value as $L$ increases (see Appendix~B and Fig.~B1). This separation between unchanged GOE-like local correlations and a strongly temperature-dependent global density distinguishes our phenomenon from Rosenzweig--Porter-type scenarios \cite{rosenzweig1960repulsion}, where the emphasis is on a crossover of local spectral statistics and eigenvector structure rather than a deformation of the global bulk density.

Thus, the observed deformation of the spectral density is a natural companion of the emergence of two-point correlations within the overlap field $\mathcal{M'}_{ij}$, Eq.~\ref{Eq-2}. Empirically the global spectrum is described by a simple pattern {\it regardless of details of the disorder distribution} (Gaussian or bimodal), in the form of a one-parameter $q$-Gaussian interpolation with $q$ evolving from $-1$ (semicircle) to $1$ (Gaussian) around criticality.

In a broader context, our findings align with recent insights into spectral behavior in disordered many-body systems. Ref.~\cite{altland2024quantum} shows that in a “low-entropy” regime---when the effective number of free parameters is small compared to the Hilbert-space dimension---collective fluctuations can dominate, softening spectral edges relative to fully chaotic (“high-entropy”) settings.
Our mechanism is distinct but similar in spirit: in the non--self-averaging window $\beta \gtrsim \beta_c$, the ensemble spectrum $P(\lambda)=\mathbb{E}_J[P(\lambda|J)]$ averages over many disorder samples whose \emph{per-sample} spectra are relatively narrow but whose centroids and widths fluctuate strongly across disorder realizations $J$ near criticality (reflecting large sample-to-sample variations in the spatial overlap field on the slice). When these sample-to-sample shifts dominate the within-sample (thermal) spread, the disorder average effectively acts like a superposition of many slightly shifted narrow spectra, which under broad and standard mixture conditions drives the resulting bulk toward a smooth Gaussian---without implying “low entropy” for any individual sample.
In this sense, the approach to a Gaussian bulk at $T_c$ is presented as an empirical, ensemble-level consequence of mixture/averaging and does not assume any specific scenario for the low-temperature spin-glass phase (e.g., droplet or RSB).
The absence of a comparable deformation for single-replica spin matrices underscores that the effect originates from inter-replica correlations in the overlap.

 Our results provide direct evidence that, already in the paramagnetic phase, complex configurational organization and the evolution of correlations manifest themselves at the level of global spectral observables. An important challenge for theories of the EA spin glass therefore is to quantitatively account for the richly structured replica overlap spectrum and its temperature dependence.
This approach can, in principle, be applied to other classical spin models (e.g., 3D Ising and Potts). It can also be carried over to quantum critical systems by constructing analogous matrices from \emph{equal-time} correlators (i.e., as correlation fingerprints rather than energy-level spectra).

\textit{Acknowledgments}---This work was funded by the Deutsche Forschungsgemeinschaft (DFG, German Research Foundation)---under Project No.~557852701 (A.A.S.). We thank Alex Altland, John Chalker, and Shivaji Sondhi for helpful discussions. This work was also supported in part by the Deutsche Forschungsgemeinschaft through Research Unit FOR~5522 (project-id 499180199) and the Cluster of Excellence ct.qmat (EXC~2147, Project No.~390858490).

\textit{Data availability}---The data that support the findings of
this article are openly available \cite{onder_2026_18391567}.

\bibliography{references}  
\onecolumngrid
\vspace{1em}
\begin{center}
\textbf{End Matter}
\end{center}
\vspace{1em}

\twocolumngrid
\clearpage

\appendix
\renewcommand\thefigure{A\arabic{figure}}
\setcounter{figure}{0}

\begin{figure}[t!]
\centering
\includegraphics[width=1.1\columnwidth]{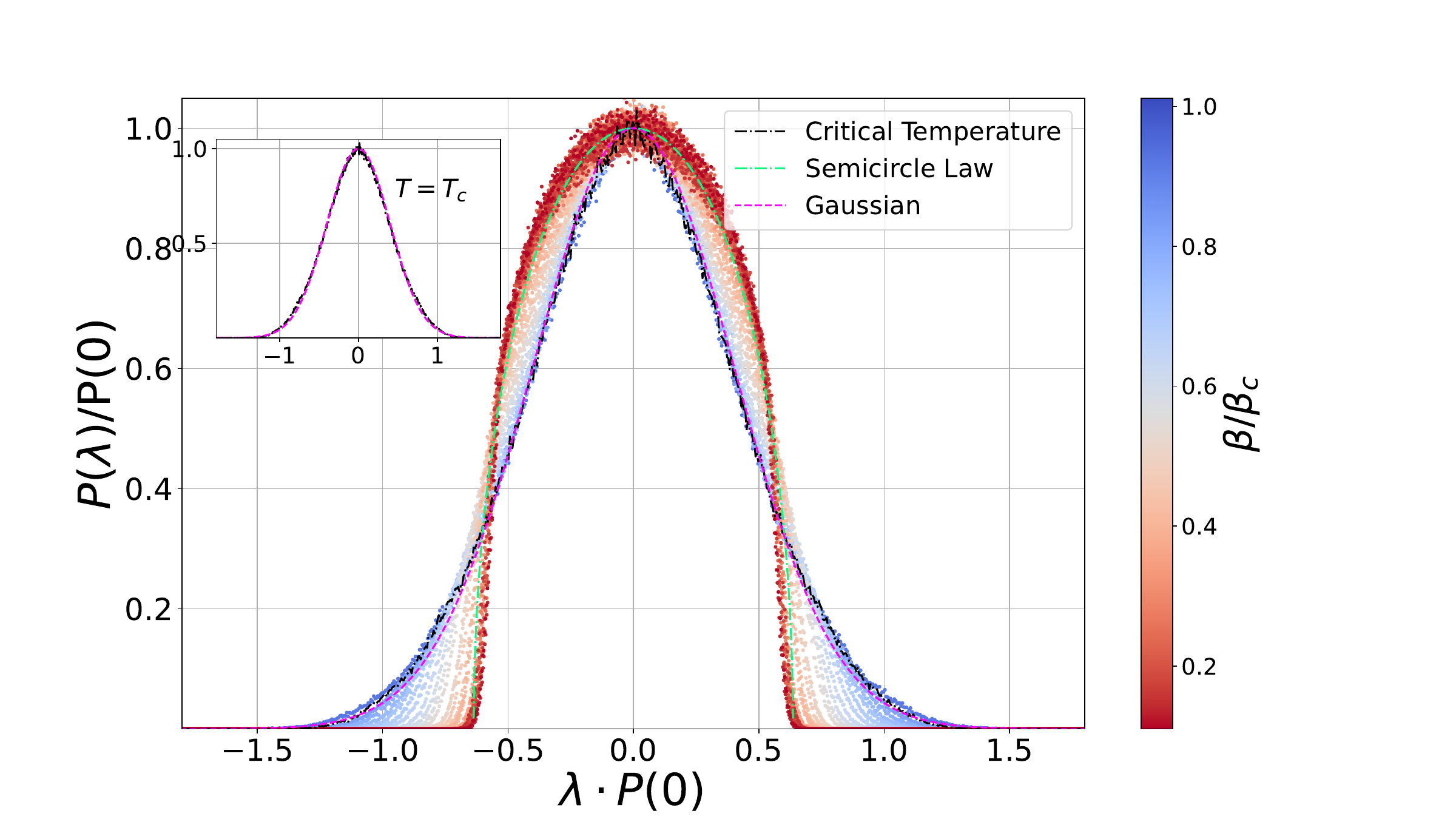}
\caption{Scaled spectral density $P(\lambda)/P(0)$ of bulk eigenvalues for the overlap matrix $\mathcal{M}$ of the 3D EA model with $\pm J$ couplings, shown for system size $L=24$ across temperatures $\beta=1/T$. A smooth crossover is observed from the Wigner semicircle (dashed green) at high $T$ to a Gaussian form (dashed magenta) near the critical point $\beta_c \approx 0.9075$ (i.e., $T_c \approx 1.102$). Inset: at $\beta=\beta_c$, the bulk density matches a Gaussian within statistical uncertainty.}
\label{fig:center30pm}
\end{figure}

\begin{figure}[t!]
\centering
\includegraphics[width=0.9\columnwidth]{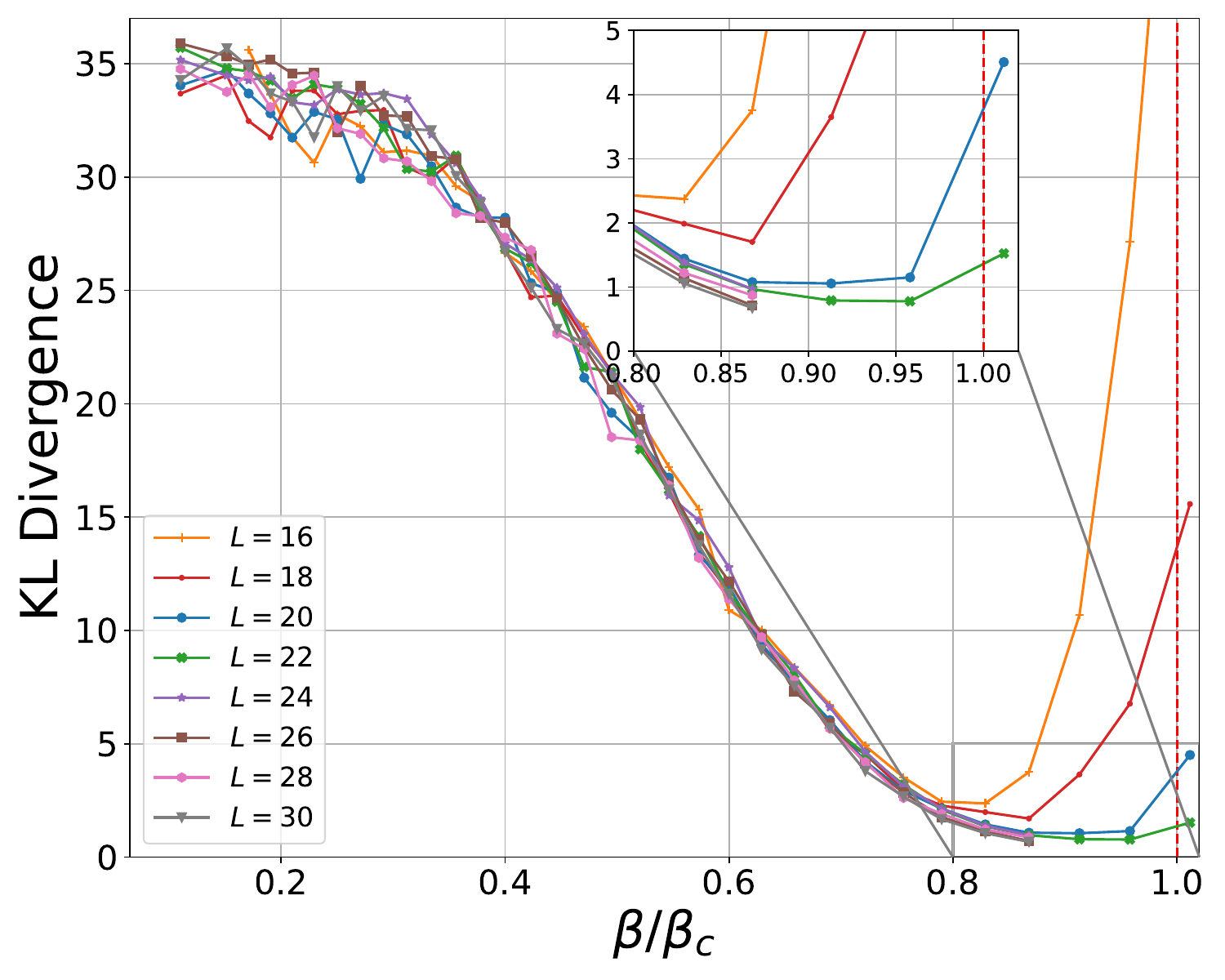}
\caption{Temperature dependence of the Kullback–Leibler divergence $D_{\rm KL}(P\!\parallel\!P_{\rm G})$ between the empirical bulk spectral density $P(\lambda)$ and a variance-matched Gaussian $P_{\rm G}(\lambda)$ for the $\pm J$ model, shown for system sizes $L=16$–$30$. For $\beta \gtrsim \beta_c \approx 0.9075$ (vertical red dashed line), $D_{\rm KL}$ decreases systematically with $L$ and tends toward zero, indicating convergence of the bulk spectral density to a Gaussian in the thermodynamic limit. Inset: zoom near criticality highlights the approach of $D_{\rm KL}$ to zero as $\beta/\beta_c \to 1$.}
\label{fig:kldiv_pm}
\end{figure}

\noindent\textit{Appendix A: Bimodal disorder ($\pm J$).}
To test robustness against the choice of disorder, we repeated the analysis for the 3D Edwards–Anderson model with $J_{\mathbf{xy}}=\pm 1$ (equal probability), using the same simulation and spectral pipelines as in the main text. Appendix Fig.~A1 shows the scaled bulk spectral density $P(\lambda)/P(0)$ for $L=24$ across temperatures: as in the Gaussian-disorder case, the spectrum exhibits a smooth crossover from the Wigner semicircle at high $T$ to a Gaussian near criticality. Quantitatively, the best agreement with a Gaussian occurs at $\beta=\beta_c\approx 0.9075$ (i.e., $T_c\approx 1.102$), where the bulk density matches a Gaussian within statistical uncertainty. Appendix Fig.~A2 reports the Kullback–Leibler divergence $D_{\rm KL}(P\!\parallel\!P_{\rm G})$ between the empirical bulk spectrum and a variance-matched Gaussian for $L=16$–$30$. For $\beta\gtrsim\beta_c$, $D_{\rm KL}$ decreases systematically with $L$ and tends toward zero, indicating that the bulk spectral density converges to a Gaussian in the thermodynamic limit. Taken together, Figs.~A1–A2 demonstrate that the semicircle-to-Gaussian crossover and the Gaussian match at criticality are insensitive to whether the couplings are Gaussian or bimodal, underscoring the disorder-agnostic character of the observed spectral phenomenology.

\renewcommand{\theequation}{B\arabic{equation}}
\setcounter{equation}{0}

\noindent\textit{Appendix B: Adjacent-gap ratio.}
To complement the global density analysis in the main text, we also probed local spectral correlations via the adjacent-gap ratio
\begin{equation}
r=\left\langle \frac{\min(\Delta_i,\Delta_{i+1})}{\max(\Delta_i,\Delta_{i+1})}\right\rangle ,
\end{equation}
where $\{\lambda_i\}$ are the bulk eigenvalues sorted in ascending order and
$\Delta_i=\lambda_{i+1}-\lambda_i$ are consecutive gaps (computed in the bulk, excluding the outlier).
Appendix Fig.~B1 shows $r$ versus $\beta/\beta_c$ for several system sizes.
Within statistical accuracy, $r$ displays no systematic temperature dependence across the range studied.
With increasing $L$, $r$ drifts upward and is expected to converge to the GOE benchmark $\langle r\rangle\simeq 0.536$,
remaining clearly separated from the Poisson value $\langle r\rangle\simeq 0.386$.
This supports the picture that temperature mainly reshapes the \emph{global} spectral density, while local level repulsion stays GOE-like.

\renewcommand{\thefigure}{B\arabic{figure}}
\setcounter{figure}{0}

\begin{figure}[t!]
  \centering
  \includegraphics[width=1.0\linewidth]{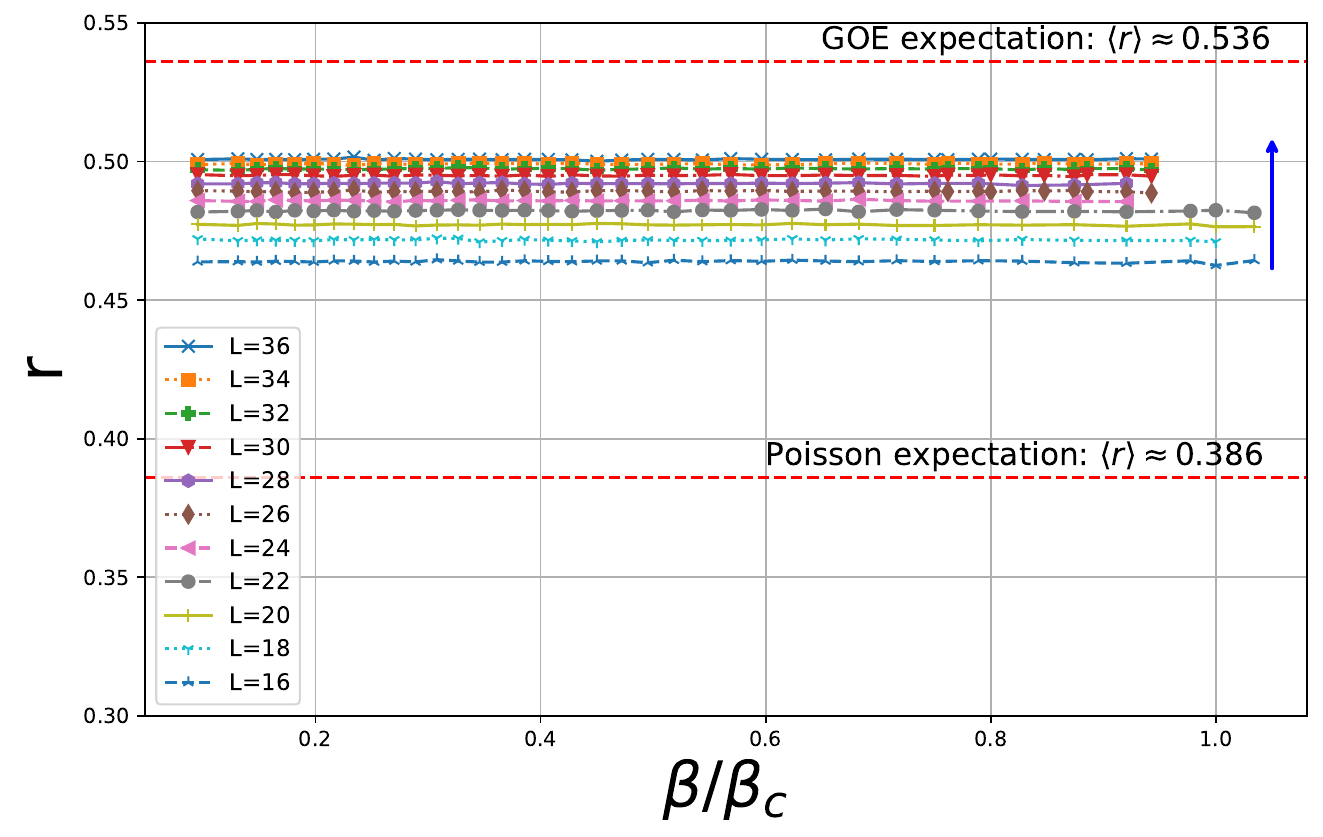}
  \caption{Average adjacent-gap ratio in the bulk as a function of $\beta/\beta_c$ for several system sizes $L$.
  The values are essentially temperature-independent over the range shown.
  The arrow indicates increasing $L$: as $L$ grows, the ratio drifts upward toward the GOE benchmark (upper dashed line),
  while remaining clearly separated from the Poisson limit (lower dashed line).}
  \label{fig:spacing_ratio}
\end{figure}

\end{document}